\documentclass[aps]{revtex4}

\newcommand {\ket}[1]{|\,{#1}\,\rangle}

\newcommand {\bra}[1]{\langle\,{#1}\,|}

\newcommand{\beq}{\begin{equation}}
\newcommand{\eeq}{\end{equation}}
\newcommand{\eq}[1]{Eq. (\ref{#1})}

\usepackage[dvips]{graphicx}

\begin{document}
\title{\bf Classical-Wigner Phase Space Approximation to Cumulative
Matrix Elements in Coherent Control}
\author{B. R. McQuarrie}
\affiliation{Division of Science and Mathematics\\University of Minnesota at Morris,
Morris, MN, USA 56267}
\author{Dmitri G. Abrashkevich and Paul Brumer}
\affiliation{Chemical Physics Theory Group\\
Department of Chemistry \\
University of Toronto, \\
Toronto M5S 3H6,\ \ Canada}

\vspace{2 cm}
\begin{abstract}
The classical limit of  the Wigner-Weyl representation is used to
approximate products of bound-continuum matrix elements that are
fundamental to many coherent control computations. The range of
utility of the method is quantified through an examination of
model problems, single-channel Na$_2$ dissociation and
multi-arrangement channel photodissociation of CH$_2$IBr. Very
good agreement with the exact quantum results is found for a wide
range of system parameters.
\end{abstract}

\maketitle

\section{Introduction}
\indent{~~} Recent experimental and theoretical developments show
that coherent control, i.e. the control of atomic and molecular
dynamics via quantum interference, can be successfully applied to
the wide variety of systems \cite{cc}. However, available quantum
methods limit theoretical studies to scattering and
photodissociation  of small molecules. The extension of
theoretical studies to larger complex chemical systems  must rely
on new developments in quantum or semiclassical techniques.
Primary amongst these is the application of the Initial Value
Representation semiclassical approach that is described elsewhere
\cite{batista,Miller1}. However, the ultimate utility of such techniques
requires the development of numerical tools to speed convergence
of the initial value integrals with highly oscillatory integrands.

In this paper we consider an approach to computing interference
contributions in coherent control in which such oscillations do not occur.
Specifically, we focus attention on using the classical limit of  Wigner
Phase Space Methods\cite{Wigner,Imre}, where the desired transition matrix
elements products are written in the Wigner-Weyl representation and the
quantum expressions are replaced by their classical counterparts.  This
approach  has been applied in the past to a number of simpler
problems\cite{Heller1,Heller3} where transitions from one initial bound
state were considered, e.g., from a single vibrational state
\cite{Heller1} or from a Gaussian state \cite{Hupper,Hupper1}. In these
cases only absolute values squared of matrix elements were sought. By
contrast, in this paper we use a result \cite{Wilkie} on the classical
limit of the nonstationary Liouville eigenstates to obtain the classical
limit of the desired product of transition dipole matrix elements, a
quantity which, in general, includes phase information. For example, we
demonstrate that the method gives good results for complex valued
matrix elements that arise in the multi-arrangement photodissociation of
CH$_2$IBr to CH$_2$I + Br or CH$_2$Br + I. The method has also been previously
applied in the guise of the linearized approximation to the
Initial Value Representation\cite{Miller2} with varying degrees of success.

This paper is organized as follows. The classical limit of the
Wigner phase space method is described in Section II. Applications
of the method to simple systems involving one arrangement channel
are described in Section III. Specifically, we consider
Franck-Condon transitions for the model case of excitations from a
harmonic oscillator potential to a linear potential and to
transitions on realistic Na$_2$ potential energy surfaces. Section
IV discusses applications to collinear photodissociation of
CH$_2$IBr, where the transitions probabilities are complex because
the relevant operators are no longer Hermitian. Section V provides
a summary of results.

\vskip 0.1truecm
\section{Method}
\indent{~~}

In various coherent control scenarios (e.g., bichromatic control
\cite{ccold} or coherent control via pulse  sequencing
\cite{Seideman}) control is dictated (here written for the case of
non-rotating diatomics) by interference terms of the form
\beq
\sigma_{n,m}^{(r)}(E) = \sum_\textbf{k} \bra{E,\textbf{k},r^-}
\mu_{fi}\vert n\rangle\langle m\vert\mu_{fi}\ket{E,\textbf{k},r^-}
= \sum_\textbf{k} \bra{E,\textbf{k},r^-}
\chi_n\rangle\langle\chi_m\ket{E,\textbf{k},r^-}~.
\label{e1}
\eeq
Here $\vert\chi_j\rangle = \mu_{fi} \vert j\rangle,\,j=n,m$ and
$\mu_{fi}$ is the electronic transition dipole moment between the
upper  and lower electronic states.  The states $\vert j\rangle$
of energy $E{_j}$, are  bound vibrational states on the lower
potential energy surface, and $\ket{E,\textbf{k},r^-}$ are
continuum nuclear eigenstates of energy $E$ on an excited
electronic surface, corresponding to product in channel $r$ with
quantum numbers \textbf{k}.
 Terms  like those in \eq{e1}  arise
when one  calculates the probability $P(E)$ of transition from a
coherently prepared bound superposition state
\beq
\vert\psi\rangle = \sum_i c_i \vert i\rangle,
\label{e2}
\eeq
to the final continuum states at energy $E$ in arrangement $r$
\cite{ccold}. That is, these terms correspond to the interference
terms in computing the probability
 $P(E) \propto
\sum_k \vert \langle E,\textbf{k},r^-\vert \mu_{fi} \vert
\psi\rangle\vert^2$.  Since coherent control relies upon quantum
interference to obtain control over molecular outcomes, the
evaluation of terms like those in \eq{e1} are vital to
computational control studies.

In the simplest case, when $n=m$ and only one product arrangement
channel is open, \eq{e1} is proportional to the photodissociation
probability for the transition from an initial bound state $\vert
n \rangle $ at energy $E_n$ to the continuum of final states at
energy $E$:
\beq
\sigma_{n,n}^{(r)}(E) = \sum_\textbf{k} \vert
\bra{E,\textbf{k},r^-}\mu_{fi}\vert n\rangle\vert^2.
\label{e3}
\eeq

To utilize \eq{e1} we rewrite it as:
\beq
\sigma_{n,m}^{(r)}(E)= {\rm Tr}[
(\sum_\textbf{k}\ket{E,\textbf{k},r^-}\bra{E,\textbf{k},r^-}
(\vert \chi_m\rangle\langle \chi_n\vert)],
\label{e4}
\eeq
where  ``Tr" denotes the trace. The term $\sum_\textbf{k}
\ket{E,\textbf{k},r^-}\bra{E,\textbf{K},r^-}$ can then be written
as
\beq
\sum_\textbf{k} \ket{E,\textbf{k},r^-}\bra{E,\textbf{k},r^-} =
\hat R_r \delta({E-\hat H})
\label{project}\eeq
where $\hat R_r$ projects onto product arrangement $r$ and $\hat
H$ is the Hamiltonian of the excited electronic state
\cite{WBO82}. When summed over $r$ (or when there is only one
product arrangement channel)  this equation reduces to the
familiar expression:
\beq
\sum_{\textbf{k},r} \ket{E,\textbf{k},r^-}\bra{E,\textbf{k},r^-} =
\delta(E-\hat H). \eeq

Using \eq{project},  \eq{e4} can then be rewritten as

\beq
\sigma_{n,m}^{(r)}(E) = {\rm Tr}[ \hat R_r\delta(E-\hat H)] (\vert
\chi_m\rangle\langle
 \chi_n\vert)].
\label{e5}
\eeq

The Wigner transform O$_W$ of any operator $\hat{\textrm{O}}$ is
defined as \cite{Wigner}
\beq
O_{\rm W} ({\bf p},{\bf q}) = {1\over \pi\hbar}
\int_{-\infty}^{+\infty} d{\bf v}\,\,\, e^{2i{\bf p}\cdot{\bf
v}/\hbar} \langle {\bf q}-{\bf v}\vert \hat{O}\vert {\bf q}+{\bf
v}\rangle ,
\label{e7}
\eeq
where ${\bf p}$ is the momentum conjugate to coordinate ${\bf q}$.
 Taking the Wigner transform of
\eq{e5} and using ${\rm Tr}(AB) = { \rm Tr}(A_{\rm W}B_{\rm W})$
gives
\beq
\sigma_{n,m}^{(r)}(E) = {\rm Tr}\left[[\hat R_r \delta(E-\hat
H)]_{\rm W} \rho^{\chi}_{n,m}\right]~,
\label{e6}
\eeq
where $\rho^\chi_{n,m}({\bf p},{\bf q})\equiv [\vert
\chi_n\rangle\langle \chi_m\vert]_{\rm W}$ is the Wigner transform
of $\vert \chi_n\rangle\langle \chi_m\vert$. Neglecting the
coordinate dependence of $\mu_{fi}$ in accord with the
Franck-Condon approximation gives
\beq
\sigma_{n,m}^{(r)} (E) = \mu_{fi}^2 {\rm Tr}\left[[\hat R_r
\delta(E-\hat H)]_{\rm W} \rho_{n,m}\right]~,
\label{e8}
\eeq
where $\rho_{n,m} = \rho_{n,m}(\textbf{p},\textbf{q}) =
[\ket{n}\bra{m}]_W$.  Equation (\ref{e8}) is exact.  In Section III
of this paper we consider applications of approximations to
\eq{e8} to cases with a single product arrangement channel.
Section IV considers the case of multiple chemical products.

\section{Single Product Arrangement Channel}

In the case where there is only a single product arrangement
channel, \eq{e8} becomes
\beq
\sigma_{n,m}(E) =   \mu_{fi}^2 {\rm Tr}[ \delta(E-\hat H)_{\rm W}
\rho_{n,m}] = \mu_{fi}^2 \int d{\bf p}\, d{\bf q}\,\,
\delta(E-\hat H)_{\rm W}\, \rho_{n,m}({\bf p},{\bf q})~.
\label{e8a}
\eeq
Here
\beq
\rho_{n,m}({\bf p},{\bf q})=[\vert  n\rangle\langle m\vert]_{\rm
W} = {1\over \pi\hbar} \int d{\bf v} \,\,\exp(2i{\bf p}\cdot{\bf
v}/\hbar) \langle {\bf q}-{\bf v}\vert n\rangle \langle m \vert
{\bf q}+{\bf v}\rangle~,
\label{e9}
\eeq
so that   $\sigma_{n,m}(E)$  appears as the overlap of two phase
space densities, one corresponding to the density on the lower
surface [$\rho_{n,m}({\bf p},{\bf q})$] and one to the continuum
density $\delta(E-\hat H)_{\rm W}$ on the upper excited surface.
Neither of these densities need be classical since they can be
negative or complex.

The classical Wigner approximation to $\sigma_{n,m}(E)$ is
obtained by taking the classical limit ($\hbar \rightarrow 0$) of
\eq{e8a}, where the classical limit of $\delta(E-\hat H)_{\rm W}$
arises by either expanding the density of states in powers of
$\hbar$, or by using the statistical operator
\cite{Wigner,Heller1} $\hat P = \exp(-\beta\hat H)$, $\beta =
1/kT$,  or via an exponentiated $\hbar$ expansion \cite{Mayer}, or
by expanding $\delta(E-\hat H)_{\rm W}$ around the identity
operator $\hat I$ times the classical Hamiltonian, $H({\bf p},{\bf
q})\cdot \hat I$ \cite{Hupper}. The lowest order term in this
expansion is $\delta(E-H({\bf p},{\bf q}))$, where $H({\bf p},{\bf
q})$ is the classical Hamiltonian associated with the upper
potential energy surface $V({\bf q})$. For a particle of reduced
mass $\mu$ in one dimension, which we focus on in this section:
\beq
H(p,q) = \frac{p^2}{2\mu} + V(q)~,
\label{e10}
\eeq
so that the lowest order classical Wigner phase space method
approximation to $\sigma_{n,m}(E)$ [denoted $\sigma_{n,m}^{c}(E)$]
 is
\beq
\sigma_{n,m}^{\rm c}(E) =
\mu_{fi}^2\int \,dp\,dq\,\, \delta(E-H(p,q)) \rho_{n,m}^{c}(p,q)
\label{e11}
\eeq
\beq
= \mu_{fi}^2\int \,dp\,dq\,\, {\delta(q-q(p,E))\over
\left\vert \left({\partial V \over
\partial q}\right)_{q=q(p,E)}\right\vert} \rho_{n,m}^{c}(p,q) = \mu_{fi}^2
\int \,dp\,{\rho_{n,m}^{c}(q(p,E),p)\over \left\vert \left({\partial V \over
\partial q}\right)_{q=q(p,E)}\right\vert},
\label{e12}
\eeq
where  $E  -  p^2/2\mu  -V(q(p,E))  = 0$, and where $\rho_{n,m}^{c}(p,q)$ is the
classical limit of $\rho_{n,m}$ .

The  form  of  $\rho_{n,m}^{c}$ depends on whether the system is
integrable or non-integrable   \cite{Wilkie}.   For   integrable
systems [\eq{e14},   \eq{e15}, Ref. \cite{Wilkie}]
\beq
\rho_{n,m}^{c}={1\over 2\pi} \delta(I(p,q) - \bar I_{n,m})
\exp[i(n-m)\theta(p,q)],
\label{e13}
\eeq
where [$I(p,q),\theta(p,q)$] are action-angle variables,
$\bar I_{n,m} = (I_n+I_m)/2$, and $I_n$ is the semiclassical action associated
with state $\vert n\rangle$ (i.e., $I_n = (n+\gamma)\hbar$, where $\gamma$ is
the  Maslov  index). A similar analytic expression is not possible for chaotic
systems \cite{Wilkie}.

In  the  cases  studied below we focus on transitions  from low
lying vibrational states of diatomics. In this case we can
approximate the potential by   an   harmonic   oscillator   of
frequency   $\omega$   and  approximate $\rho_{n,m}^{c}(p,q)$  by
the   harmonic case. However, in the harmonic case the classical
$\rho_{n,m}^{c}(p,q)$  can be chosen as equal to $\rho_{n,m}(p,q)$
for the quantum  harmonic oscillator \cite{JaffeBrumer1}.
Substituting expressions for the harmonic oscillator states
\beq
\langle q \vert n \rangle = \left({\mu \omega \over \hbar}\right)^{1/4}
\left( 2^n n!\sqrt{\pi}\right)^{-1/2} H_n\left(\sqrt{\mu \omega \over \hbar} q
\right) e^{-{\mu \omega q^2\over 2\hbar}}
\label{e14}
\eeq
into the Wigner transform of \eq{e9}, we obtain
$$
\rho_{n,m}^c (p,q) =
\rho_{n,m}(p,q) = {(-1)^n \over \pi \hbar} \left[ {2^n m! \over n! 2^m}
\right]^{1/2} \left({1\over\hbar\mu\omega}\right)^{{n-m\over2}}
$$
\beq
\times \quad
 \left[ ip-\mu\omega q\right]^{n-m} \exp\left({-{2I(p,q)\over \hbar}}\right)
L_m^{n-m}\left[ {4I(p,q)\over\hbar}\right],
\label{e15}
\eeq
where $n\geq m$,   $L_k^\alpha$ is the
generalized  Laguerre  polynomial,  and  the classical action $I(p,q)$ for the
harmonic oscillator is
\beq
I(p,q) = {H(p,q) \over\omega} = {1\over 2} \left( {p^2\over \mu\omega} +
\mu\omega q^2\right).
\label{e16}
\eeq
For the case of $n=m$, \eq{e15} takes the well-known form
\beq
\rho_{n,m}(p,q) = {(-1)^n\over \pi\hbar} \exp\left({-{2H(p,q)\over
\hbar\omega}} \right) L_n\left( {4H(p,q)\over\hbar\omega}\right).
\label{e17}
\eeq

\section{Results:  Single Arrangement Channel}

\subsection{Model Potentials}

We  first test the utility of this approximation on a simple
standard model: excitation from an  harmonic oscillator initial
state potential to a linear, repulsive excited state potential
$V(q) = -\beta q + E_0$, with $E_0$ arbitrary and $\beta>0$. This
model was previously examined for $n=m=0$ in Ref. \cite{Heller1},
and can be used to  approximate transitions to an arbitrary
potential if $\beta$ is taken as the slope of the upper potential
energy surface at the peak of the initial state ($n=m=0$)
wavefunction. For simplicity, we set the $\mu_{fi}$ to unity.
$\delta (E-\hat{H})_W$ is known \cite{Heller1} for the linear
excited state potential, so that the exact $\sigma_{n,m}(E)$ is
given by
$$
\sigma_{n,m}(E) = 2\pi\int \, dp\,dq\, \delta(E-\hat H)_W
\rho_{n,m}(p,q)
$$
\beq
=  2\pi\int \, dp\,dq\, \left({\beta^{1/3}\over\pi\beta}\right)
Ai\left[ -(2\beta^{1/3}) (q-q(p,E))\right] \,\rho_{n,m} (p,q).
\label{e18}
\eeq
where $Ai$ is the Airy function.

By contrast, the classical result [\eq{e12}] for this model is
$$
\sigma_{n,m}^{\rm c}(E) = \int \, dp\,dq\,
\delta\left(E-{p^2\over2\mu} + \beta q -E_0\right)
\rho_{n,m}^{c}(p,q)
$$
\beq
= {1\over\beta}\int \, dp\,dq\, \delta(q-q(p,E)) \,\rho_{n,m}^{c}
(p,q) = {1\over\beta}\int \, dp\, \rho_{n,m}^{c}(p,q(P,E)),
\label{e19}
\eeq
where  $q(p,E)  =  (E_0-E)/\beta + p^2/2\mu\beta$.

  In  Fig. 1 we compare the exact quantum and classical Wigner results
    for the highly quantum
case  of $\mu=m_e=$1 a.u., where $m_e$ is the mass of electron.
Note that $\sigma_{n,m}$, $n\neq m$ is real for the
one-dimensional case with real $V(q)$ since the integral over the
imaginary part is odd in the momentum variable. In Figs. 1(a)-(b),
$\sigma_{n,m}$ for $n=m=0$ and $n=m=4$ are shown as functions of
energy $E-E_0$ for parameters given in Ref. \cite{Heller1}; $\beta
= 6$ a.u., $\omega = 2$ a.u. Our results for $n=0$ agree very well
with those in Ref. \cite{Heller1}, and with the $n=m=1$ results
for $\beta=6$ a.u.,  $\omega=1$ a.u. (not shown). However, as is
evident from Fig. 1, the accuracy of the classical Wigner
approximation deteriorates extremely rapidly with increasing $n$;
results are very poor even for $n=4$. The same behavior is
observed for cases where $n\neq m$, shown in Fig. 1(c)-(d) for
($n=1,\,m=0$) and ($n=5,\,m=4$). This is because for small $n$,
$\rho_{n,n}$ is smooth and broad, and the transition integral
averages over many of Airy function oscillations. By contrast, for
large $n$, $\rho_{n,n}$ is highly oscillatory and the initial
state probes fine details of the Airy functions which are absent
in the classical approximation. As is evident from Fig. 1(b), the
classical Wigner results for $n=m$ are negative at some points.
This is impossible physically and indicative of errors in the
approximation. In the $n\neq m$ case, where negative values are
possible  [Fig. 1(c)-(d)], the positions of maxima in the
semiclassical approximations are shifted to lower energies,  a
feature previously explained for the $n=m$ case
\cite{Heller1,Hupper1}.

\begin{figure}[ht]
\begin{center}
\rotatebox{90}{\includegraphics[width = 0.55 \textwidth]{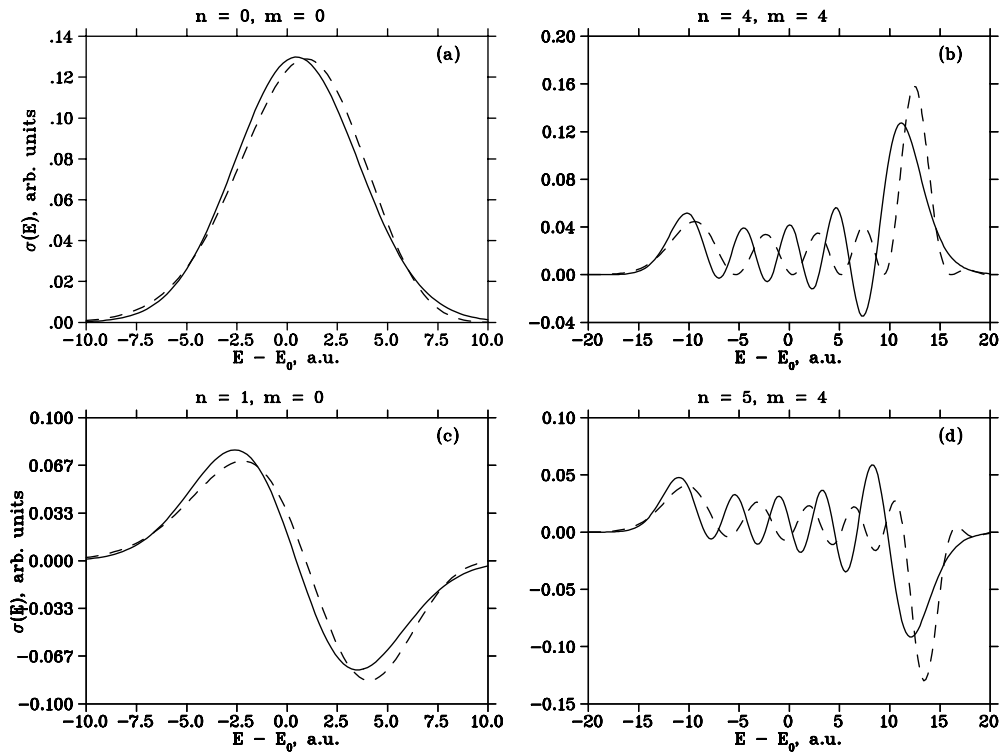}}
\end{center}
\caption{Comparison between the classical Wigner (solid line) and exact
quantum (dashed line) results for $\sigma_{n,m}(E-E_0)$ $(n=m)$ for the
$\mu = $ 1 a.u. and $\omega=2$ a.u.:
(a) $n=m = 0$; (b) $n = m = 4$;
(c) $n=1, m = 0$; (d) $n = 5, m = 4$.}
\end{figure}

Results presented in Fig. 1 were obtained for $\mu=1$ a.u. which
is highly non-classical system. The dependence of the
classical-quantum agreement on  the  oscillator frequency $\omega$
and on the reduced mass $\mu$ are explored, for  two cases with
$m=n$, in Fig. 2. Agreement is seen to improve dramatically as
$\omega$ decreases and somewhat as $\mu$ decreases.  The latter
result is surprising, motivating the analysis described below.

\begin{figure}[ht]
\begin{center}
\rotatebox{90}{\includegraphics[width = 0.55 \textwidth]{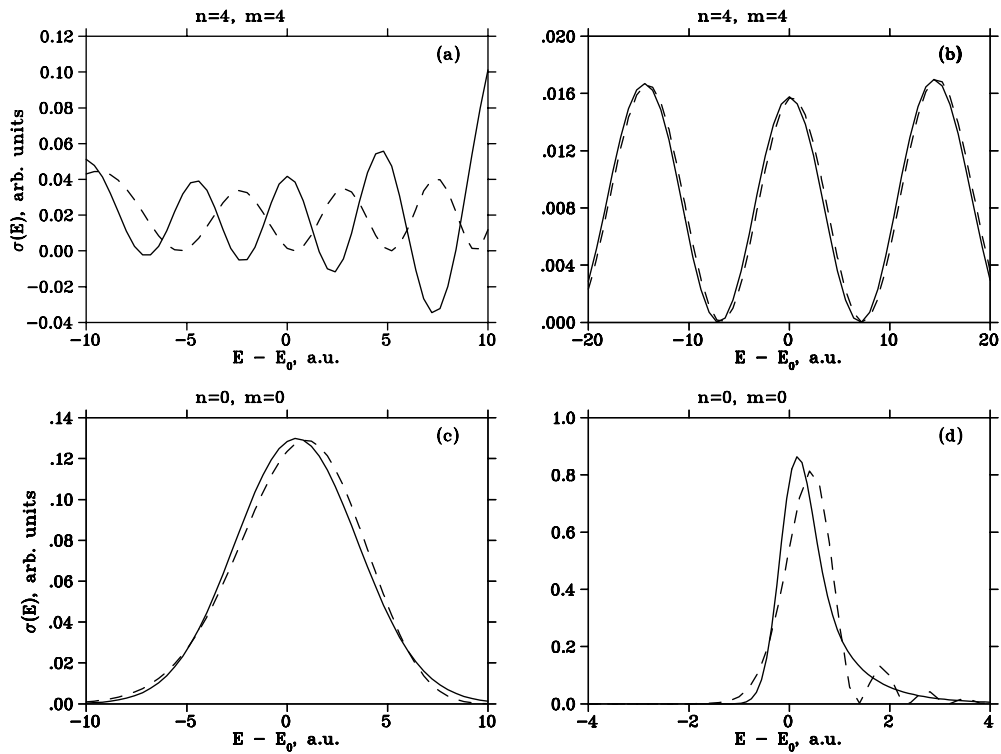}}
\end{center}
\caption{Comparison between the classical Wigner (solid line) and exact
quantum (dashed line) results for $\sigma_{n,m}(E-E_0)$ $(n=m=4)$
for fixed $\mu=$1 a.u.: (a) $\omega= 2.0$; (b) $\omega=0.2$; and
for fixed $\omega=$2 a.u.: (c) $\mu= 1.0$; (d) $\mu=100.0$.}
\end{figure}

{\em Characteristic $s_n$ factor}. The validity of the classical
Wigner  approximation  for  cases where $n=m$ and for the excited
state linear potential has been  examined  in  Ref. \cite{Hupper}.
There, the utility  of the Wigner phase space approximation for an
initial Gaussian wavefunction was shown to depend upon two
parameters, $s$ and $\lambda$:
\beq
s={\lambda\over
\Delta}, \qquad \lambda = \left( {\hbar^2\over 2\mu\beta}\right)^{1/3},
\label{add1}
\eeq
where $\lambda$ sets the scale for the width of the excited state
wavefunction's oscillations near the turning point (other
oscillations have shorter wavelength), and $\Delta$ is the width
of the initial Gaussian ground state:
\beq
\langle q\vert\Psi_{\rm gr}\rangle = {1 \over \pi^{1/4}\Delta^{1/2}}
e^{-(q-q_0)^2/2\Delta^2}.
\label{add2}
\eeq
Specifically, Eckhardt and coworkers \cite{Hupper1} have shown
that the smaller the $s$, the more accurate the classical
approximation, consistent with the fact that smaller $s$ means
more Airy oscillations over the width $\Delta$. However, the
parameter $s$, as defined in \eq{add1}, is not as useful in our
case since we consider assorted ground state vibrational wave
functions, and not just simple Gaussians. Specifically, $s$ was
unable to predict the correct dependence of the classical-quantum
agreement on $\mu$ or $\omega$. This is because for any
oscillator, the width of the vibrational state $\Delta$ depends on
both $\mu$ and $\omega$, whereas \eq{add1} only allows for a
dependence  on $\mu$ via $\lambda$.

To generalize the $s$ expression to higher vibrational states,
for cases with $m = n$,  we compare the known expression for the $v=0$
vibrational level
\beq
\psi=\left( {\alpha\over\pi} \right)^{1\over 4}e^{-\alpha(q-q_0)^2\over 2},
\quad \alpha = {\omega \mu \over \hbar},
\label{add3}
\eeq
to \eq{add2}, and obtain $\Delta = {1\over \sqrt{\alpha}}=\sqrt{\hbar\over
\mu \omega}$. The $s$ parameter can therefore be written as
\beq
s = {\lambda\over \Delta} = \left( {\hbar\mu\omega^3\over 4 \beta^2}
\right)^{1/6}.
\label{add4}
\eeq
Further, to account for the oscillatory character of the $n$th
wavefunction on the lower potential surface,  $\Delta$ is replaced
by $\Delta_n$, where $\Delta_n$ equals the width at half-maximum
of a single oscillation of the ground state wavefunction. This is
given by $\Delta_n = l/2(n+1)$, where
$l=2\sqrt{2(n+1/2)\hbar/\mu\omega}$ is the overall width of the
ground state wavefunction, estimated as the distance between the
two classical turning points. This gives the new parameter $s_n$:
\beq
s_n = {\lambda\over \Delta_n} = \sqrt{(n+1)^2\over
2(n+1/2)}\,\,\,s \;\; ; \; {\rm or} \; \;  s_n = \sqrt{n\over
2}\,\, s {\rm \,\,\,\,for\,\,large\,\,}n.
\label{add5}
\eeq
It follows from \eq{add5}  that the smaller the $s$ is for
$n=m=0$, the  better  the  classical  approximation for
higher-lying levels. However,  $s_{n}$  will  always  increase
with  increasing  $n$  due  to  the decreasing wavelength of the
bound wavefunction. Further, decreasing the vibrational frequency
of the lower electronic state $\omega$ leads to  wider vibrational
states (and larger $\Delta$ in \eq{add1} for the Gaussian ground
state) and, therefore, to the smaller  $s$ and better agreement
with quantum result. This explains the results shown in Fig.
2(a)-(b), where the corresponding values of  $s$ for two $\omega$
values  are 0.62 ($s_4=1.03$) and 0.20 ($s_4=0.3$), respectively.
Alternatively, note that for smaller $\omega$, the density
$\rho_{n,m}$ becomes wider in $q$ and the overlap integral in
\eq{e8a} averages over more Airy function oscillations. Increasing
the slope of the upper potential also leads to better agreement
between the classical Wigner and quantum results in accord with
\eq{add4}, reflecting the fact that the Airy function oscillates
more  with increasing $\beta$ [\eq{e18}]. Further, \eq{add4} and
\eq{add5} quantify the dependence on reduced mass $\mu$ that is
evident in Fig. 2(c)-(d) where corresponding values of parameter
$s$ for the two increasing values of $\mu$ are 0.62 and 1.33,
respectively. This is because, while $\lambda$ in \eq{add1}
decreases with increasing $\mu$, the width of the initial state
$\Delta$ does so as well.

{\em Characteristic $s_{n,m}$ factor}. In the case of $n\neq m$,
developing  the corresponding parameter $s_{n,m}$ is  more
difficult. However, in accord with $s_n$, the quantity $s_{n,m}$
is expected to be the ratio of a characteristic width on the
excited state divided by a width on the ground state. To obtain
$s_{n,m}$ we make the following observations: (1) The width $l_m
=2\sqrt{2(m+1/2)\hbar/\mu\omega}$ of the product $\phi_n \phi_m$
can be defined by the width of the narrower of the two states $m$,
with $m<n$,
 where $\phi_n$ is given by
\eq{e14}, since $\phi_m \rightarrow 0$ outside that interval; (2)
the total number of relevant nodes of the product $\phi_n \phi_m$
is $n'+m'$, where $n'$ and $m'$ are the number of zeroes of the
$n$th and $m$th states within the interval $l_m$.
Further, the quantity $m'=m$ since
$l_m$ is the overall
width of the $m$th harmonic oscillator state and $n'$ can be estimated as
$l_m/\Delta_n$, where $\Delta_n$ is the characteristic width of $\phi_n$:
\beq
\Delta_n = {l_n\over n+1} = {2\sqrt{2(n+1/2)\hbar/\mu\omega}\over n+1}.
\eeq
Hence, the number $N$ of oscillations of $\phi_n \phi_m$ is given by
\beq
N={l_m\over \Delta_n} + m + 1.
\eeq
Thus, the characteristic width at half-maximum $\Delta_{n,m}$
of the oscillations of the
product $\phi_n \phi_m$
is
\beq
\Delta_{n,m} = {l_m/2\over N} = {l_m/2\over {l_m\over \Delta_n} + m + 1}=
{\sqrt{2(m+1/2)\hbar/\mu\omega} \over (n+1)\sqrt{2m+1\over 2n+1} + m + 1}
\eeq
The parameter $s_{n,m}$ is then
\beq
s_{n,m}={\lambda\over \Delta_{n,m}} = {\sqrt{2m+1\over 2n+1} (n+1)
+m+1 \over
 (m+1)} s_m.
\eeq
Substituting
\beq
s_m = {\lambda\over \Delta_m} = {m+1\over \sqrt{2m+1}}{s\over 2},
\eeq
we obtain the pleasing result that \beq s_{n,m} = 1/2 \left(s_n +
s_m\right). \label{threeone}\eeq As expected, for $n=m$, $s_{n,n}
= s_n$.

The behavior of the results  in Fig. 1  can now be quantified in
terms of $s_n$ and $s_{n,m}$. In particular, we obtain the
following values of $s_n$ and $s_{n,m}$  for the results presented
in the figures: $s_0\equiv s=0.62$, $s_4=1.03$, $s_{1,0}=0.67$,
and $s_{5,4}=1.08$. One can see that $s$ is already larger than 1
(fast diverging series; see Table 1 in \cite{Hupper})  for $n=4$.
Additional computations show that $s_{4,0}$ = 0.83 and the
approximation works reasonably well in this case. Similar good
agreement has been obtained for three different pairs of $n$ and
$m$ which are characterized by $s_{n,m}$  $\approx$ 0.83, e.g.,
$(n,m)$ = (4,0), (3,1), and (2,2). However, in all cases where
$s_{n,m} > 1$ (e.g., $\approx$1.12 for $(n,m)$ = (8,2), (7,3), and
(5,5)), the agreement is poor (not shown).

\subsection{Molecules}

To apply this to realistic systems, consider first a linear
potential model of the dissociation of the H$_2$ molecule. In this
case $\mu$ is increased  and the ground state $\omega$ is
decreased relative to the system studied above. In particular,
$\mu$ = 918.7 a.u. and $\omega = \omega_{e}$ = 4395.2 cm$^{-1}$
\cite{Herzberg}. $\sigma$ results for four pairs of $n$ and $m$
are shown in Fig. 3 for $\beta$ = 6: $n=m=4$, $n=m=20$, $n=5,m=4$,
and $n=21, m=12$. The results clearly show much better agreement
between exact Franck-Condon and classical results than does the
case studied above. Indeed, the difference between the classical
and exact quantum results becomes visible only at high-lying
levels, $n=m=20$. Furthermore, this difference becomes practically
indiscernible for the even heavier molecule Na$_{2}$, $\mu$ =
20953.9 a.u. $\omega = \omega_{{\rm Na}_{2}}=$ 158.91 cm$^{-1}$
\cite{Schmidt},  where virtually perfect agreement between exact
and approximate results is obtained even for $n=20$ (not shown).
Using \eq{add4} we obtain the corresponding values of parameter
$s$: $s_{[5]} = 0.62$ (i.e., the parameters are those of Ref.
\cite{Heller1} and used in Section IV.A above), $s_{\rm H_{2}} =
0.19$, and $s_{\rm Na_{2}} = 0.06$, in accord with the computed
results. Similarly, the excellent agreement obtained for the high
$n$ and $m$ values for model H$_2$ and Na$_2$ is consistent with
the values of $s_n$ and $s_{n,m}$. For example, $s_{20}$= 0.197
for Na$_2$ and $s_{21,12}$ = 0.179. In the case of the H$_2$ model
 $s_{20}$ = 0.62 and we expect to see the difference between the
classical and exact quantum results on a par with the one we have
seen for the case in Section IV.A where $s_0\equiv s$ was also
0.62. This is indeed the case, as seen from a comparison of Figs.
1 and 3. Note also that $s_{21,12}$ = 0.57 in the case of the
H$_2$ model which leads to worse agreement than in the Na$_2$ case
($s_{21,12}=0.18$) but much better  than in the  case in Section
IV.A ($s_{21,12}=1.85$) (not shown here).

\begin{figure}[ht]
\begin{center}
\rotatebox{90}{\includegraphics[width = 0.55 \textwidth]{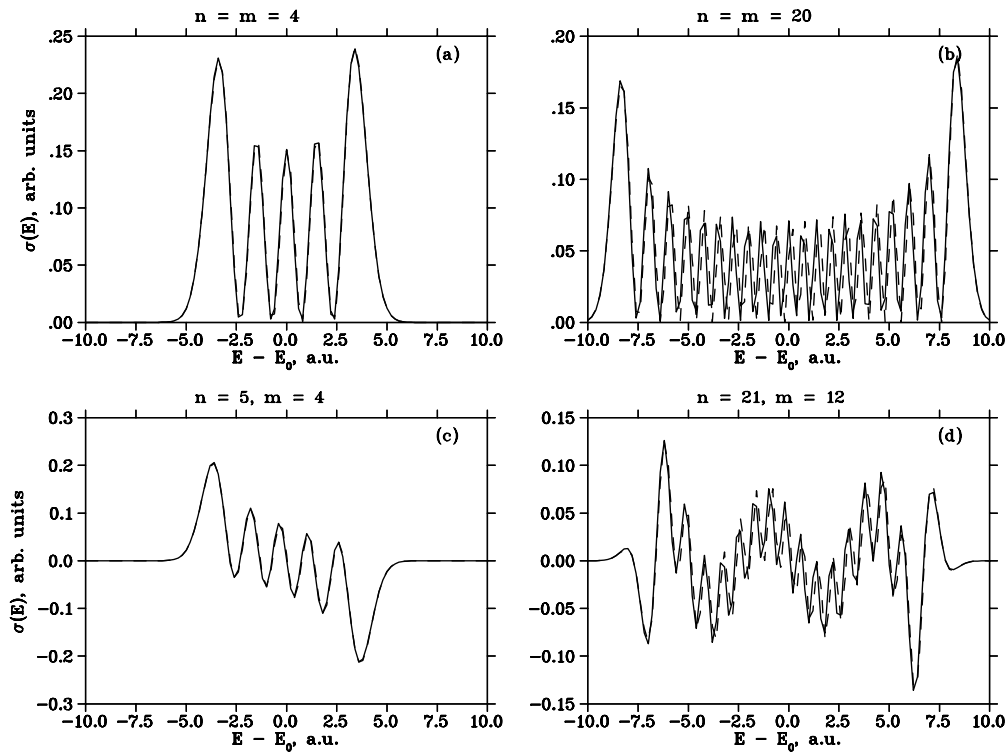}}
\end{center}
\caption{Comparison between the classical Wigner (solid line) and exact
quantum (dashed line) results for $\sigma_{n,m}(E-E_0)$ $(n=m)$
with $\mu = \mu_{\rm H_{2}}$ and $\omega =
\omega_{\rm H_{2}} =$ 4395.2 cm$^{-1}$:
(a) $n = m = 4$; (b) $n = m = 20$;
(c) $n = 5, m = 4$; (d) $n = 21, 12$.}
\end{figure}

To see the origin of this behavior, consider cuts through the
ground and excited phase space distributions at fixed values $p=0$
and $E=E_0$, as shown in Fig. 4 for three different systems: the
model of Ref. \cite{Heller1}, H$_{2}$ and Na$_{2}$. While the
quantum results correspond to taking the overlap between these two
phase space densities, the classical limit corresponds to using
the value of the initial state phase space distribution at the
coordinate $q(p,E)$ (the vertical line in Fig. 4 corresponds to
the value $q(p=0,E=E_0)=0$ for $p=0,E=E_{D}$). Clearly, this
approximation improves with decreasing $s$.

\begin{figure}[ht]
\begin{center}
\includegraphics[width = 0.45 \textwidth]{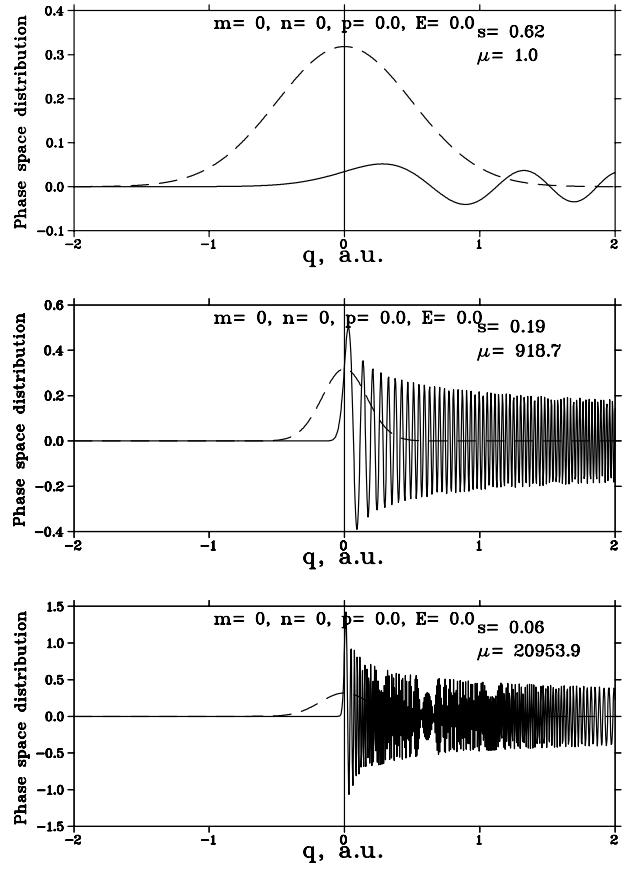}
\end{center}
\caption{Phase space densities $\rho_{5,4}$ (dashed line) and
$\delta(E-H)_{\rm W}$ (solid line) at fixed $p=0$ and $E=E_0$
for three different systems. Vertical line corresponds to the classical limit
$\delta(q-q(p,E))$.}
\end{figure}

In an attempt to further improve  these  results, we considered
the higher order  quantum corrections (see Eq. (4.8), Ref.
\cite{Heller1}) to the classical Wigner result, which consist of
additional terms in an expansion of the density of states
$\delta(E-\hat H)_{\rm W}$ in powers of $\hbar$. Our calculations
using this expansion showed that although these corrections
sometimes work well for the low-lying vibrational levels $n < 2$,
they led to much poorer agreement with the exact results
for higher $n$, as the higher order terms became dominant, a
result also noted in Ref. \cite{Hupper}. As is evident from Fig.
5, where the quantum corrected results are shown as dot-slashed
curves, the quantum corrections are practically of no use when
$n,m$ are $>0$ even for a small value of $s$ ($s_{\rm
Na_{2}}=0.06$). The results are equally  bad for the other
models and are not shown here.

\begin{figure}[ht]
\begin{center}
\rotatebox{90}{\includegraphics[width = 0.25 \textwidth]{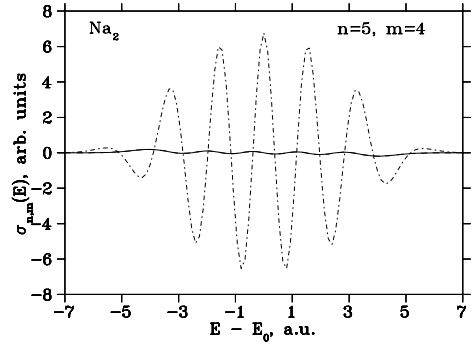}}
\end{center}
\caption{Comparison between the classical Wigner and exact quantum (solid line)
and quantum corrected Wigner (dot-dashed line) results
for the $n=5, m=4$,
$\mu = \mu_{\rm Na_{2}}$ and $\omega = $ 158.91 cm$^{-1}$.}
\end{figure}

Consider now the case of a realistic  model  of  Na$_2$
\cite{Schmidt} where  both  the upper and lower potentials are
properly treated; the  ${\rm b} \,1 {}^3\Pi_u$ to $1 ^3\Pi_g$
transition has been chosen as an example. The results
using the classical Wigner approximation (solid line)
and the uniform semiclassical approach (dashed line) are compared
in Fig. 6(a)-(f). Here the classical calculations were carried out
using \eq{e12} with realistic Na$_2$ potentials, whereas the
semiclassical Franck-Condon factors  were obtained by numerically
evaluating
\beq
\langle E^- \vert j\rangle = \int\,dq\, \langle E^- \vert q\rangle
\langle q\vert j\rangle, \quad j=n,m,
\label{e20}
\eeq
using Gauss-Legendre quadrature. Here, $q$ is the Na-Na
separation, $\langle q\vert j\rangle$ are the bound vibrational
wave functions of the initial electronic state calculated quantum
mechanically using the Renormalized Numerov method \cite{Johnson},
and $\langle q\vert E^-\rangle$ are the continuum wave functions
of the excited potential energy surface, calculated using the
uniform semiclassical approximation \cite{Langer1},\cite{Miller}
\beq
\langle q\vert E^-\rangle={\xi (q)^{1\over4} \over K(q)^{1\over
2}} Ai(-\xi(q))e^{-i\delta}.
\label{e21}
\eeq
where
\beq
\xi(q)=\left[{3\over 2}\int_a^qdq^{\prime} K(q^{\prime})
\right]^{2/3},
\label{e22}
\eeq
with
\beq
K(q)=p(q)/\hbar,  \quad p(q)=\{2\mu(E-V(q))\}^{1\over 2},
\label{e23}
\eeq
\beq
\delta = \int_a^\infty [K(q)-k]\, dq - k a + {\pi\over 4},
\quad k = \lim_{q\rightarrow \infty} K(q),
\label{e24}
\eeq
and $a$ is the classical turning point, which satisfies
$V(q=a)=E$. As can be seen in Fig 6 the results clearly indicate
that the classical Wigner representation works very well for the
low-lying levels ($n,m\leq 5$) of the initial electronic state.
The results for very high $n$ and $m$ are in poorer agreement and
are not presented here. This is a direct consequence of the use of
the harmonic approximation for the ground state. Specifically, for
$n,m>5$, anharmonic corrections to the ground state should be
included in order to account for the delocalization of the ground
state wavefunctions \cite{Heller4}.

\begin{figure}[ht]
\begin{center}
\rotatebox{90}{\includegraphics[width = 0.55 \textwidth]{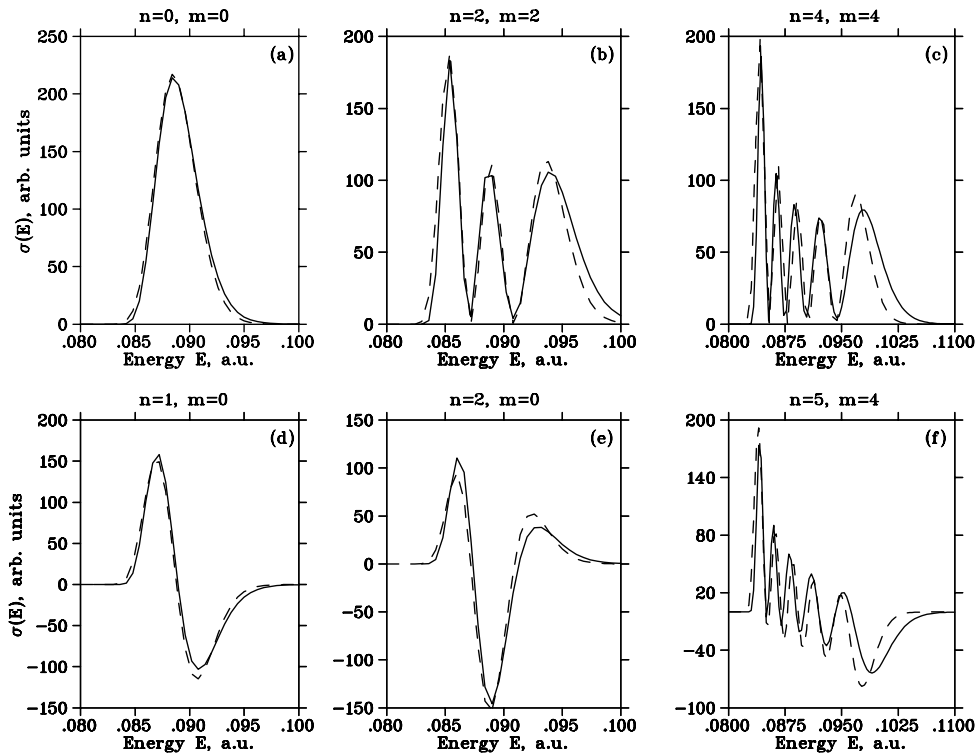}}
\end{center}
\caption{Comparison between the classical Wigner (solid line) and the uniform
semiclassical approximation (dashed line) calculations carried out for the
realistic Na$_2$ potentials:
(a) $n=m = 0$; (b) $n = m = 2$; (c) $n = m = 4$;
(d) $n=1, m = 0$; (e) $n = 2, m = 0$; (f) $n = 5, m = 4$.}
\end{figure}

\section{Multi-Product Arrangement Channels}

Consider now the multi-arrangement channel problem, i.e., the case
where photodissociation results in the formation of two different
chemical products, e.g., $A + BC \leftarrow ABC \rightarrow AB +
C$.  In this case our main focus is on obtaining cross sections
into specific channels.

Quantum mechanically the Hilbert space of a typical
multi-arrangement channel scattering problem can be partitioned as
follows~\cite{WBO82}:
\begin{equation}
\hat{I} = \sum_r \hat{R}_r + \hat{B} .
\end{equation}
Here, $\hat{I}$ is the unit operator, $\hat{R}_r$ projects onto states
that correlate asymptotically with all states in channel $r$, and
$\hat{B}$ projects onto bound states. This allows the
total $\sigma_{n,m}(E)$ to be written as
\begin{eqnarray}
\sigma_{n,m}(E) &=& \sum_r {\rm Tr} \left[\hat{R}_r \delta (E-\hat{H}) |n
\rangle\langle m |\right]
+ {\rm Tr} \left[\hat{B} \delta (E-\hat{H}) |n \rangle\langle m |\right], \nonumber \\
&=& \sum_r \sigma_{n,m}^{(r)}(E) + \sigma_{n,m}^{(B)}(E) .
\end{eqnarray}
The channel-specific cross section of interest in this section is
given by \eq{e5} in the Franck-Condon approximation, i.e.,
\begin{equation}
\sigma_{n,m}^{(r)}(E) = \mu_{fi}^2 \textrm{Tr}[\hat R_r \delta(E-\hat H)
\ket{n} \bra{m}]
\label{chans}
\end{equation}

Classically, these  operators $R_r$ and $B$ correspond to various
types of classical trajectories that occur in photodissociation:
trajectories that start in the region of the excited polyatomic
(upon excitation) and dissociate to  the  $r$ channels, and those
that do not dissociate and remain bound.

Equation (\ref{chans}), in the Wigner representation, assumes the
form
\begin{eqnarray}
\sigma_{n,m}^{(r)}(E)& =& {\rm Tr} \left[ [\hat R_r \delta(E-\hat
H)]_W [\ket{n}\bra{m}]_W \right] \nonumber \\ & \equiv & \int dz [
\hat R_r \delta(E-\hat H)]_W(z) \rho_{n,m}(z)
\end{eqnarray}
where $z$ denotes all coordinates and momenta, $q,p$.

To evaluate the channel-specific cross section requires that we
 approximate the term $[\hat{R}_r \delta
(E-\hat{H})]_W$.  In general, the Wigner transform of the product
of two operators admits a small $\hbar$ expansion~\cite{Gro46}
\begin{equation}
(\hat{A}\hat{B})_W(z) = \hat{A}_W(z)\hat{B}_W(z) + \frac{i\hbar}{2}
\left\{ A_W , B_W \right\}_{\rm p}(z) + O(\hbar^2),
\end{equation}
where $\left\{\cdot,\cdot\right\}_{\rm p}$ is the Poisson bracket.
Therefore, the channel specific cross section can be approximated by
\begin{eqnarray}
\sigma_{n,m}^{(r)}(E) &\sim& \int dz \: \rho_{n,m}(z) \label{PBV1} \\
&\mbox{}\times& \left[ [\hat{R}_r]_W(z) \delta (E-\hat{H})_W(z)
+\frac{i\hbar}{2} \left\{ [\hat{R}_r]_W , \delta (E-\hat{H})_W
\right\}_{\rm p}(z) \right]. \nonumber
\end{eqnarray}
Equation~(\ref{PBV1}) can be rewritten by employing the cyclic invariance
of the trace, so that
\begin{eqnarray}
\sigma_{n,m}^{(r)}(E) &=& \int dz\: [\hat{R}_r]_W(z) \label{CSRHO1} \\
&\times& \left[ \delta (E-\hat{H})_W(z)\rho_{n,m}(z) + \frac{i\hbar}{2}
\left\{ \delta (E-\hat{H})_W,\rho_{n,m} \right\}_{\rm p}(z) \right].
\nonumber
\end{eqnarray}
This form is more natural  since the term that selects the channel,
$[\hat{R}_r]_W(z)$,  acts in the same manner on both terms in the
integral.

We can implement this channel selection as follows: we consider
the trajectory that emanates from an initial point $z$; if the
trajectory ends in channel $r$ it contributes to
$\sigma_{n,m}^{(r)}(E)$.  Alternatively, it contributes to
channel $r' \neq r$, or is bound, both of which are ignored in the
$\sigma_{n,m}^{(r)}$ computation.

Both the magnitude and phase of $\sigma_{n,m}^{(r)}(E)$ are
important in coherent control. Hence we note that this term  can
have an imaginary part if $n\neq m$, since the integration over
momentum is now constrained and arguments based on the odd vs.
even nature of the integrand do not apply. To see this more
clearly, consider $\sigma_{n,m}^{(r)}(E) = \langle m |
\hat{R}_r\delta(E-\hat{H}) | n \rangle$. The wavefunctions $| j
\rangle$ are real, and therefore if $\sigma_{n,m}^{(r)}(E)$ is to
have an imaginary component, the operator
$\hat{R}_r\delta(E-\hat{H})$ must be non-Hermitian. Since each of
$\hat{R}_r$~\cite{WBO82} and $\delta(E-\hat{H})$ are individually
Hermitian, the operator $\hat{R}_r \delta(E-\hat{H})$ is
non-Hermitian only if $[\hat{R}_r,\delta(E-\hat{H})] \neq 0$. This
is indeed the case, as can be seen by taking the matrix element
with respect to eigenfunctions of the Hamiltonian $\hat{H}$, to
find that
\begin{eqnarray}
\langle E^{'} | [\hat{R}_r,\delta(E-\hat{H})] | E^{''} \rangle &=& \langle
E^{'} | \hat{R}_r\delta(E-\hat{H}) | E^{''} \rangle -
\langle E^{'} | \delta(E-\hat{H})\hat{R}_r | E^{''} \rangle ,\nonumber \\
&=& \langle E^{'} |\hat{R}_r| E^{''} \rangle (\delta(E-E^{''}) -
     \delta(E-E^{'})) \neq 0 .
\end{eqnarray}
However, even though $\sigma_{n,m}^{(r)}(E)$ can have an imaginary
component, the sum of the imaginary contributions over all the
channels must remain zero since $\sigma_{n,m}(E)$ is real. For a
two channel system, this implies that any imaginary contribution
in channel 1 must be equal in magnitude and opposite in sign to
the imaginary contribution from channel 2.  We further note that
the diagonal $\sigma_{n,n}^{(r)}(E)$ is
\begin{eqnarray}
\sigma_{n,n}^{(r)}(E) &=& \langle n|\hat{R}_r\delta(E-\hat{H}) |n
\rangle,
\nonumber \\
&=& \sum_\textbf{k} \bra{n} E,\textbf{k},r^- \rangle\langle E,\textbf{k},r^-\ket{n},\nonumber \\
&=& \sum_\textbf{k} \vert \bra{n} E,\textbf{k},r^-\rangle\vert^2
\nonumber
\end{eqnarray}
and is therefore real.

Consider now the second term in Eq.~(\ref{CSRHO1}). The
derivatives of the delta function introduced via the Poisson
bracket result in the rapid oscillation of the integrand which, as we
have verified numerically, yields a final contribution to the
integral that is essentially zero. Therefore, we set the Poisson
bracket term to zero, and the channel specific $\sigma_{n,m}^{(r)}(E)$
becomes
\begin{equation}
\sigma_{n,m}^{(r)}(E) \sim \int dz\: [\hat{R}_r]_W(z) \delta
(E-\hat{H})_W(z) \rho_{n,m}(z). \label{CSRHO}
\end{equation}

Equation (\ref{CSRHO}) has the form of two overlapping phase space
densities. Specifically, the density $\rho_{n,m}(z)$ overlaps
$[\hat{R}_{r}]_W \delta (E-\hat{H})_W$, where the latter is the phase
space density of states on the energy shell at energy $E$ that decays to
product in channel $r$.   To successfully approximate \eq{CSRHO} requires
a classical approximation to the Wigner transform $[\hat{R}_r]_W [\delta
(E-\hat{H})]_W$, discussed below.

\subsection{Computational Results}

The lowest dimensionality problem of this kind, useful for
examining the utility of the semiclassical approximation under
consideration, is the collinear photodissociation of ABC as $A +
BC \leftarrow ABC \rightarrow AB +C $, where each of $A$, $B$, $C$
denote an atom or molecular fragment. We consider this problem
below, where the electronic ground state is assumed to be well
approximated by an harmonic oscillator.

For the two degrees of freedom case we use the notation $z =
(p_1,p_2,q_1,q_2)$, $p = (p_1,p_2)$, $q = (q_1,q_2)$, $n =
(n_1,n_2)$, $m = (m_1,m_2)$, where $p_i, q_i$ denote the momenta
and coordinates of the two degrees of freedom. The two dimensional
harmonic oscillator initial vibrational state is given by
\begin{equation}
 \langle q_1|n_1\rangle\langle q_2|n_2\rangle=
N_{n_1} H_{n_1}(\sqrt{\alpha_1} q_1) e^{-\alpha_1 q_1^2/2} N_{n_2}
H_{n_2}(\sqrt{\alpha_2} q_2) e^{-\alpha_2 q_2^2/2},
\end{equation}
and the two dimensional Wigner function can be written as the
product of two one dimensional Wigner functions
\begin{equation}
\rho_{n,m}(z) = \rho_{n_1,n_2,m_1,m_2}(z) =
\rho_{n_1,m_1}^{(\alpha_1)}(p_1,q_1)
\rho_{n_2,m_2}^{(\alpha_2)}(p_2,q_2).
\end{equation}

The  excited state Hamiltonian is
\begin{equation}
H(z) = \frac{p_1^2}{2M_1} + \frac{p_2^2}{2M_2} + V(q_1,q_2),
\end{equation}
In the computations below it proves advantageous to numerically approximate
the delta function as
\begin{equation}
\delta (E-\hat{H})_W \sim \delta (E-H(z)) \sim
\frac{1}{2\sqrt{\pi\epsilon}} \exp \left\{ \frac{-(E-H(z))^2}{4\epsilon}
\right\},
\label{deltaf}
\end{equation}
where $\epsilon$ is chosen small. The final form for the
approximate channel specific term in two dimensions then
becomes
\begin{equation}
\sigma^{(r)}_{n_1,n_2,m_1,m_2}(E) \sim
\frac{1}{2\sqrt{\pi\epsilon}} \int dz \: [R_r(z)]_W
\rho_{n_1,n_2,m_1,m_2}(z) \exp \left\{
\frac{-(E-H(z))^2}{4\epsilon}\right\}~. \label{way1}
\end{equation}

Results were also obtained by  expanding the delta function to
reduce the dimensionality of the integrand from four to three:
\begin{equation}
\delta(E-H(z)) =
\frac{\delta(p_2-p_2^+)}{|p_2^+/M_2|}+\frac{\delta(p_2-p_2^-)}{|p_2^-/M_2|},
\end{equation}
where
\begin{equation}
p_2^{\pm} = p_2^{\pm}(q_1,q_2,p_1,E) =
\pm\sqrt{2 M_2 \left[E - p_1^2/(2 M_1) - V(q_1,q_2)\right] }.
\end{equation}
so that the channel specific term is now given by
\begin{equation}
\sigma^{(r)}_{n_1,n_2,m_1,m_2}(E) = \sum_{\tilde{p}_2}
\int_{\Gamma} dq_{1}dq_{2} dp_1 [R_r(q_1,q_2,p_1,\tilde{p}_2)]_W
\rho_{n_1,n_2,m_1,m_2}(q_1,q_2,p_1,\tilde{p}_2)
\frac{M_2}{|\tilde{p}_2|}, \label{way2}
\end{equation}
where the integral over $\Gamma$ requires that
$E \geq V(q_1,q_2) + p_1^2/(2M_1)$, and the sum is over
$\tilde{p}_2 = p_2^{\pm}$.

Monte Carlo integration of Eq.~(\ref{way1}) and (\ref{way2})
showed that Eq.~(\ref{way1}) converged with fewer trajectories,
even though the integral is of higher dimension. Further, the
ability to smooth the integral by increasing $\epsilon$ allows
qualitative estimates of the form of the channel specific cross
sections with a small number of trajectories. Some sample results
are provided below.

\section{Application to CO$_2$ and CH$_2$BrI}
\label{sec4}

The method was first applied to the photodissociation of collinear CO$_2$.
The coordinates are denoted $R = r_{\rm O-C}, r = r_{\rm C-O}$. The system
is initially in the bound state $\Psi_b (r,R)$, with equilibrium
separation $\bar{x}_s=\bar{R}=\bar{r}=2.20\;{\rm a.u.}$ The ground
potential surface is harmonic in the normal mode coordinates $x_s$ and
$x_a$~\cite{KCO91}, with parameters:
\begin{eqnarray}
&&x_s = (R+r)/2; \; x_a = (R-r)/(2\gamma); \; \gamma= [1+m_C/(2 m_O)],\\
&&M_s = 2m_O; \; M_a = m_C ( 1 + m_C/(2m_O)),\\
&&\alpha_s = \omega_s M_s/\hbar; \; \alpha_a = \omega_a M_a/\hbar.
\end{eqnarray}
The two product channels are denoted OC+O (channel 1) and O+CO
(channel 2). The coordinates $q_1 \equiv x_s, q_2 \equiv x_a$ are
related to the bond length coordinates by
\begin{eqnarray}
r-\bar{r} &=& q_1 - \gamma q_2 ,\nonumber \\
R-\bar{R} &=& q_1 + \gamma q_2 .
\end{eqnarray}
The multidimensional integrals in Eqs. (\ref{way1}) and (\ref{way2}) for
both this case as well as the case of CH$_2$IBr discussed below were
carried out for all systems below using Monte Carlo box Muller
transformation~\cite{NR96} and 10$^6$ trajectories. The value of
$\epsilon$ was chosen as $\epsilon=5 \times 10^{-6}$ for the case of
CO$_2$ and $\epsilon=1 \times 10^{-8}$ for CH$_2$IBr.

 Results for CO$_2$ are shown in Figs. 7 and 8.
Figure 7 compares the total cross section $\sigma_{0000}(E)$ calculated
using Eq.~(\ref{way1}), computed with $R_r(z)=1$,  to results of a time
dependent formalism utilizing the stationary phase Herman Kluk (SPHK)
propagator~\cite{MCQ99}. Trajectories in the SPHK procedure were followed
only long enough to capture the initial dispersion from the Franck-Condon
region, giving the direct part of the cross section. Although the
classical-Wigner result is seen to be shifted slightly from the time
dependent result, the method is seen to display the essential features of the
direct part of the photodissociation cross section. An analagous
calculation by Eckhardt and H\"{u}pper\cite{Hupper1} for water produced a
result similar to that shown in Fig. 7.

\begin{figure}[ht]
\begin{center}
\includegraphics[width = 0.35 \textwidth]{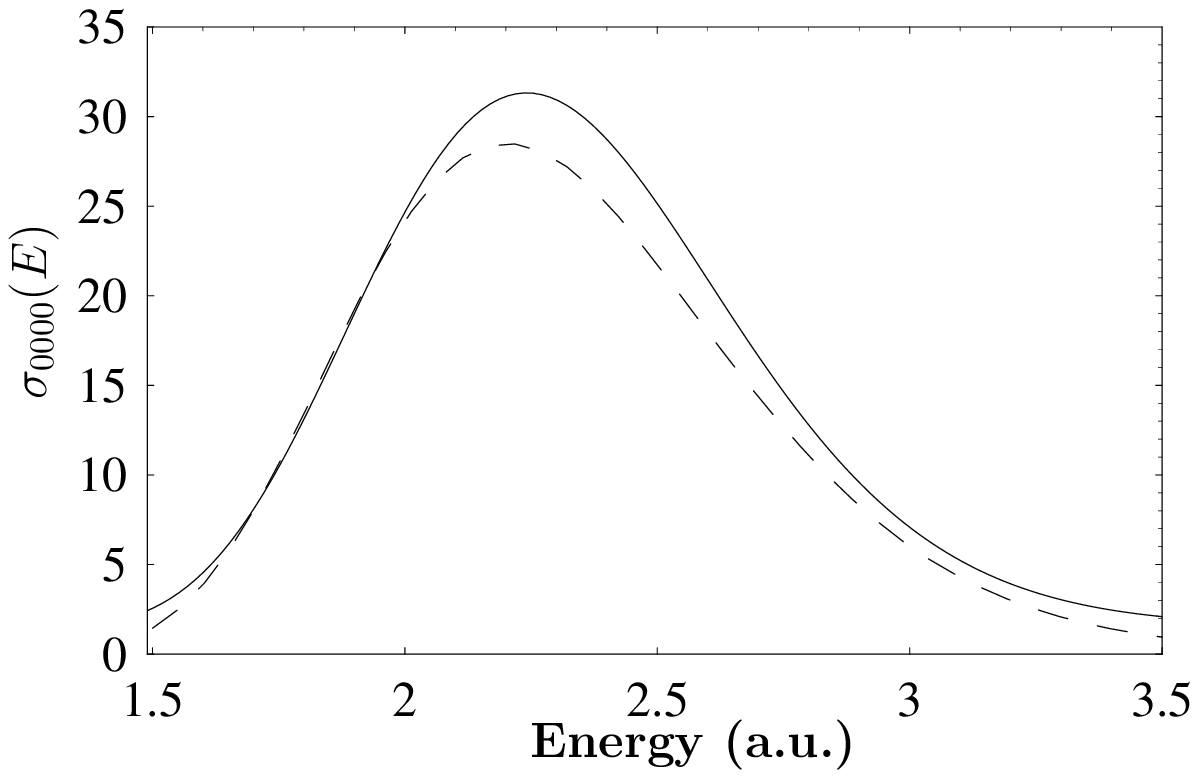}
\end{center}
\caption{CO$_2$ results for $\sigma_{0000}(E)$ (total result) using the time
dependent (autocorrelation) formalism (dash) and the time independent
(solid) formalism.}
\end{figure}

\begin{figure}[ht]
\begin{center}
\includegraphics[width = 0.65 \textwidth]{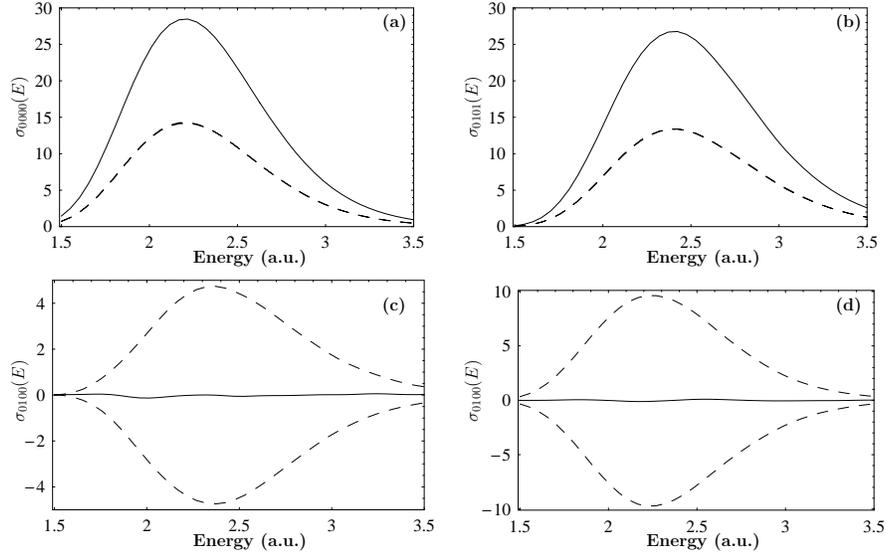}
\end{center}
\caption{CO$_2$ results. The solid line is the sum from both channels,
the dash lines are the two channel contributions.
(a) ${\rm Re}[\sigma_{0000}^{(k)}(E)]=\sigma_{0000}^{(k)}(E)$;
(b) ${\rm Re}[\sigma^{(k)}_{0101}(E)]=\sigma^{(k)}_{0101}(E)$;
(c) Re[$\sigma^{(k)}_{0100}(E)$];
(d) Im[$\sigma^{(k)}_{0100}(E)$].
}
\end{figure}

Figure 8 shows $\sigma_{0000}^{(r)}(E),\sigma_{0101}^{(r)}(E)$,
and $\sigma_{0100}^{(r)}(E)$ for CO$_2$. In this case the diagonal
cross sections are strictly real. The off-diagonal cross section
$\sigma_{0100}^{(r)}(E)$ do contain an imaginary component in both
channels, but they are seen to be equal and opposite in magnitude,
verified by computing the total cross section which is again
strictly real.

To test the utility of this approach on a system where the two channels
are dissimilar, we consider  CH$_2$BrI  where channel 1 is CH$_2$Br + I
and channel 2 is CH$_2$I + Br. The upper potential energy surface is given
by~\cite{AB01}
\begin{eqnarray}
V(r_{\rm CH_2-Br},r_{\rm CH_2-I}) &=& [V_{\rm I}(r_{\rm CH_2-I}) +
V^{\rm Morse}_{\rm CH_2-Br}(r_{\rm CH_2-Br})] f(x) + \\
&&\mbox{}[V_{\rm Br}(r_{\rm CH_2-Br}) +
V^{\rm Morse}_{\rm CH_2-I}(r_{\rm CH_2-I})] ( 1 -f(x) ) \nonumber
\end{eqnarray}
where
\begin{eqnarray}
V^{\rm Morse}_{\rm CH_2-Br}(r_{\rm CH_2-Br}) &=&
D^e_{\rm CH_2-Br} \exp \left\{ -\alpha_{\rm Br} (r_{\rm CH_2-Br} -
r_{\rm CH_2-Br}^e) \right\} \nonumber \\
&&\times [ \exp\left\{ -\alpha_{\rm Br} (r_{\rm CH_2-Br} -
r_{\rm CH_2-Br}^e) \right\} - 2] \\
V^{\rm Morse}_{\rm CH_2-I}(r_{\rm CH_2-I}) &=&
D^e_{\rm CH_2-I} \exp \left\{ -\alpha_{\rm I} (r_{\rm CH_2-I} -
r_{\rm CH_2-I}^e) \right\} \nonumber \\
&&\times [ \exp\left\{ -\alpha_{\rm I} (r_{\rm CH_2-I} -
r_{\rm CH_2-I}^e) \right\} - 2] \\
V_{\rm I}(r_{\rm CH_2-I}) &=&
A_{\rm I} \exp \left\{ -\beta_{\rm I} r_{\rm CH_2-I} \right\}, \\
V_{\rm Br}(r_{\rm CH_2-Br}) &=&
A_{\rm Br} \exp \left\{ -\beta_{\rm Br} r_{\rm CH_2-Br} \right\}, \\
f(x) &=& \frac{1}{1 + \exp \left\{ \alpha(x-0.5) \right\}} , \\
x &=& \frac{r_{\rm CH_2-Br}}{r_{\rm CH_2-I} + r_{\rm CH_2-Br}}.
\end{eqnarray}
The parameters for this surface are:
$\alpha=30$,
$D^e_{\rm CH_2-Br} = 0.1069$ a.u.,
$\alpha_{\rm Br} = 0.9154$ (a.u.)$^{-1}$,
$r_{\rm CH_2-Br}^e = 3.6850$ a.u.,
$A_{\rm Br} = 0.27$,
$\beta_{\rm Br} = 0.35$,
$D^e_{\rm CH_2-I} = 0.0874$ a.u.,
$\alpha_{\rm I} = 0.87094$ (a.u.)$^{-1}$,
$r_{\rm CH_2-I}^e = 4.04326$ a.u.,
$A_{\rm I} = 0.37$,
$\beta_{\rm I} = 0.3$.

The initial state is taken to be  harmonic in normal mode coordinates.
These coordinates, $q_1,q_2$, are related to the bond length coordinates
$r_{\rm CH_2-Br}$ and $r_{\rm CH_2-I}$ by
\begin{eqnarray}
r_{\rm CH_2-Br} - \bar{r}_{\rm CH_2-Br} &=& c_{11} q_1 + c_{21} q_2 \\
r_{\rm CH_2-I} - \bar{r}_{\rm CH_2-I} &=& c_{12} q_1 + c_{22} q_2
\end{eqnarray}
where $c_{11}=0.552747$; $c_{21}=1.09614$; $c_{12}=0.788417$;
$c_{22}=-1.03893$; $\bar{r}_{\rm CH_2-Br}=r_{\rm CH_2-Br}^e$;
$\bar{r}_{\rm CH_2-I}=r_{\rm CH_2-I}^e$.
The parameters for this system are:
\begin{eqnarray}
&&M_1 = 162614.16 \mbox{ a.u.}; \; M_2 = 27238.15 \mbox{ a.u.},\\
&&\alpha_1 = 149.3978 \mbox{ a.u.}; \; \alpha_2 = 98.8412 \mbox{ a.u.},\\
&&\hbar\omega_1 =201.638 \mbox{ cm}^{-1}; \;
\hbar\omega_2 =796.43112 \mbox{ cm}^{-1}.
\end{eqnarray}

Results for several cross sections and interference terms are shown in
Figs. 9 to 11. The diagonal cross sections are strictly real. For this
system, unlike CO$_2$, the real part of the channel specific results can
be different in the two channels.

\begin{figure}[ht]
\begin{center}
\includegraphics[width = 0.65 \textwidth]{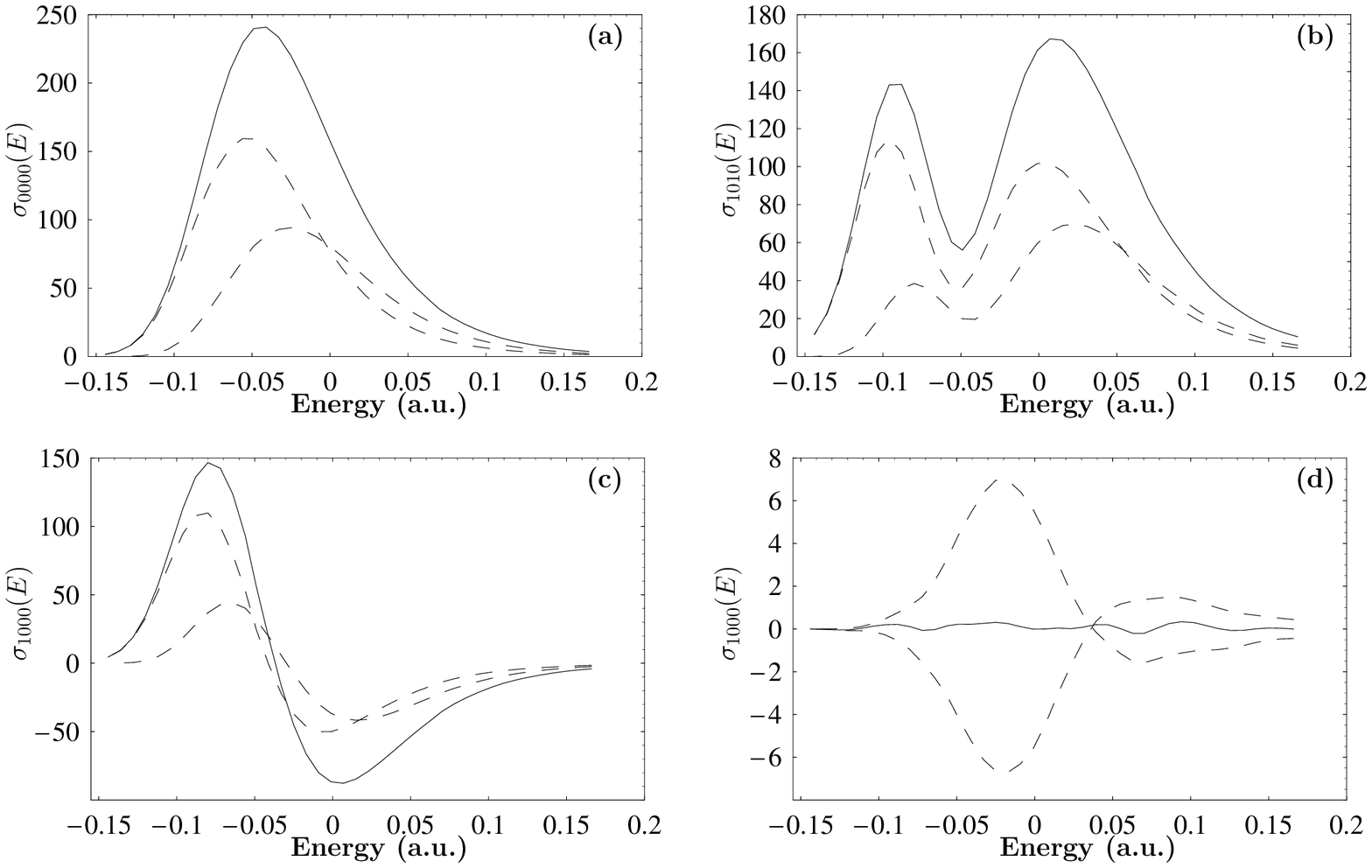}
\end{center}
\caption{CH$_2$BrI results. The solid line is the sum from both
channels, the dash lines are the two channel contributions. (a)
${\rm Re}[\sigma^{(k)}_{0000}(E)]=\sigma^{(k)}_{0000}(E)$; (b)
${\rm Re}[\sigma^{(k)}_{1010}(E)]=\sigma^{(k)}_{1010}(E)$; (c)
Re[$\sigma^{(k)}_{1000}(E)$]; (d) Im[$\sigma^{(k)}_{1000}(E)$].
}
\end{figure}

\begin{figure}[ht]
\begin{center}
\includegraphics[width = 0.65 \textwidth]{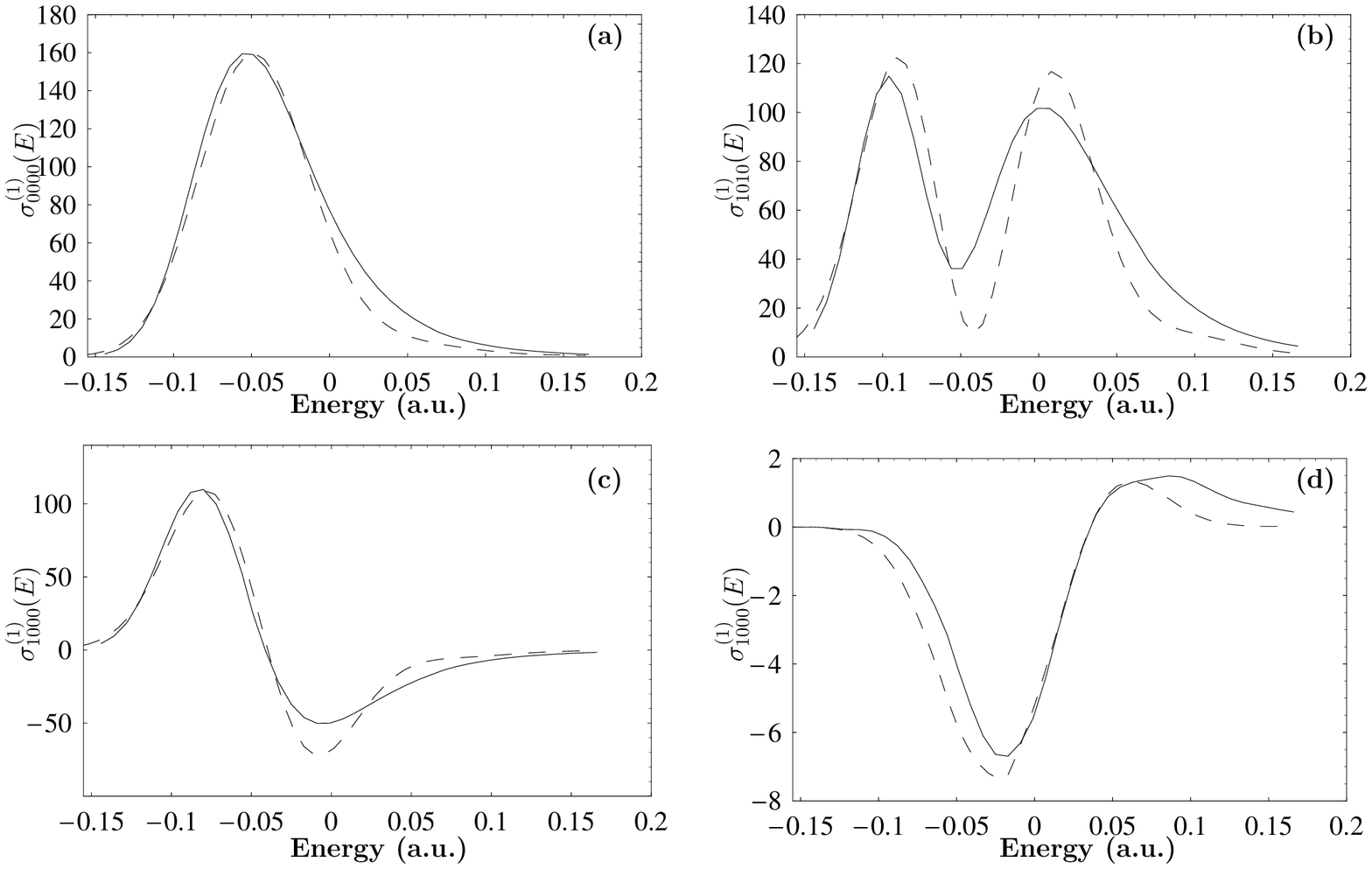}
\end{center}
\caption{CH$_2$BrI results, comparing the Wigner-classical method
(solid) to the quantum (dash) for the CH$_2$Br + I channel. (a)
${\rm Re}[\sigma^{(1)}_{0000}(E)]=\sigma^{(1)}_{0000}(E)$; (b)
${\rm Re}[\sigma^{(1)}_{1010}(E)]=\sigma^{(1)}_{0000}(E)$; (c)
Re[$\sigma^{(1)}_{1000}(E)$]; (d) Im[$\sigma^{(1)}_{1000}(E)$].
}
\end{figure}

\begin{figure}[ht]
\begin{center}
\includegraphics[width = 0.65 \textwidth]{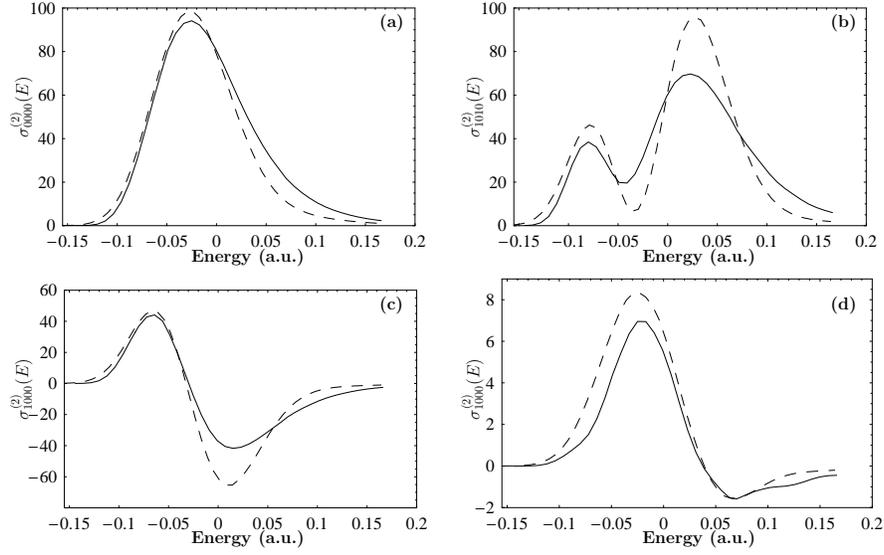}
\end{center}
\caption{CH$_2$BrI results, comparing the Wigner-classical method
(solid) to the quantum (dash) for the CH$_2$I + Br channel. (a)
${\rm Re}[\sigma^{(2)}_{0000}(E)]=\sigma^{(2)}_{0000}(E)$; (b)
${\rm Re}[\sigma^{(2)}_{1010}(E)]=\sigma^{(2)}_{1010}(E)$; (c)
Re[$\sigma^{(2)}_{1000}(E)$]; (d) Im[$\sigma^{(2)}_{1000}(E)$].}
\end{figure}

Figure 9 shows the structure of the channel contributions to various cross
sections and interference terms. Note that, as required, for the case of
the off-diagonal cross section $\sigma_{1000}^{(k)}(E)$ is interesting
(cf. Fig. 9 ), even though the two channels produce different real parts,
the imaginary parts in the two channels again sum to zero. The deviation
from zero is very small, indicating reliable convergence of the Monte
Carlo sums.

Figures 10 and 11 show a comparison of the classical-Wigner method
 with a full quantum mechanical calculation. The structure of the cross section
in the channels is seen to be well reproduced the classical-Wigner
method, providing support for the conclusion that this approach
gives reliable results for both the real and imaginary
contribution. Interestingly, the latter arises via the
non-Hermitian character of $\hat{R}_r \delta(E-\hat{H})$.

Finally, Fig. 12 shows an example of the utility of larger values
of $\epsilon$ in Eq. ({\ref{deltaf}). Specifically, using larger
values of $\epsilon$ allows for a qualitative estimate of the
desired integrals using far fewer trajectories.

\begin{figure}[ht]
\begin{center}
\includegraphics[width = 0.35 \textwidth]{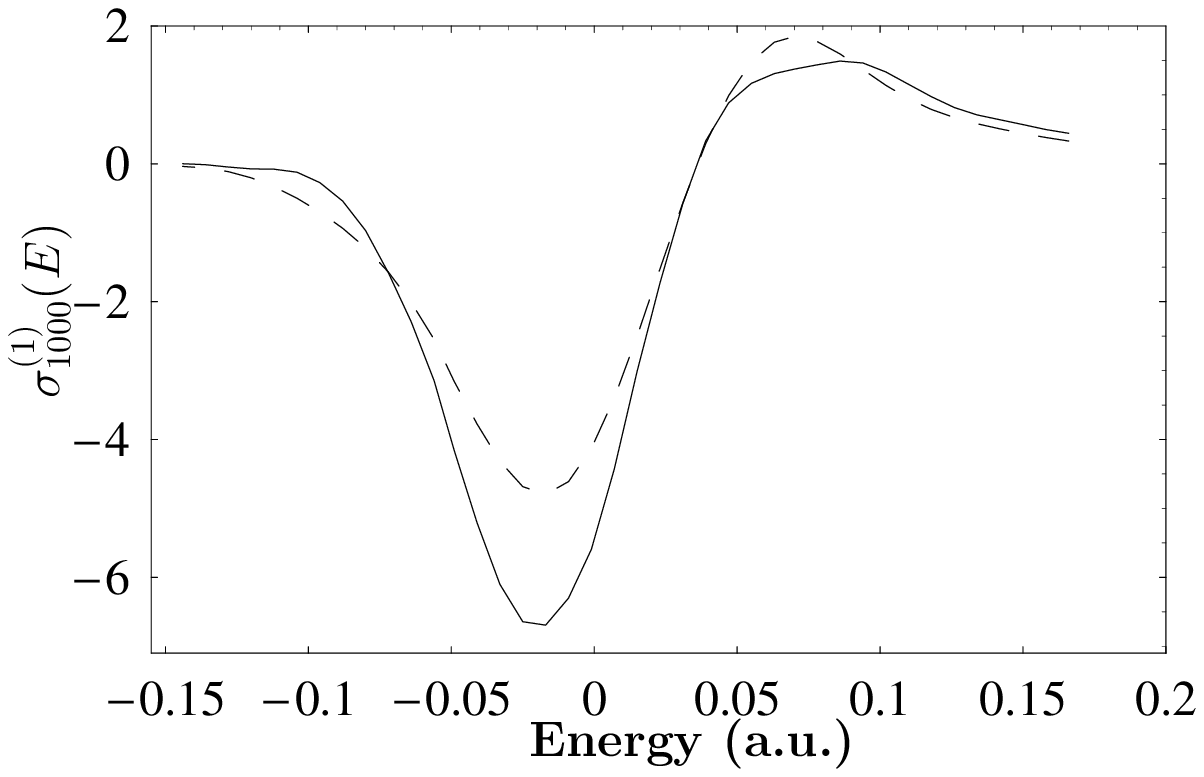}
\end{center}
\caption{CH$_2$BrI results for $\sigma_{0000}^{(1)}(E)$ (CH$_2$Br + I) with
different $\epsilon$. Dash: $\epsilon=5\times 10^{-7}$, $2\times 10^4$
trajectories; Solid: $\epsilon= 10^{-8}$, $ 10^6$ trajectories per energy.}
\end{figure}

\vskip 0.1truecm
\section{SUMMARY}
\indent{~~}

We have considered a Wigner-based classical approximation  [\eq{e11}] to
compute the terms $\sum_r \langle E,k^- \vert\mu_{fi}\vert n\rangle\langle
m\vert\mu_{fi}\vert E,k^-\rangle$, with $n\neq m$, that are central to the
interference contributions characteristic of coherent control. In the case
of a single open product arrangement channel the accuracy of this formula
for the Franck-Condon transitions onto a linear potential, along with the
dependence  of the method on parameters such as system reduced mass, slope
of the upper potential energy surface, and vibrational frequency of the
lower electronic state were examined. The results were found to be in
excellent agreement when the newly introduced parameter $s_{n,m}$
[\eq{threeone}] was less than unity.  A comparison of the classical and
uniform semiclassical approximations for the transitions between the
realistic potential energy surfaces of Na$_2$ demonstrates that higher
values of $m,n$ require use of the anharmonic Wigner function for the
ground state, an effect not explored in this paper.

This approach was also applied to the multi-channel problem where
interference terms are complex. In this case, the result is still the
overlap of two time-independent phase space densities, but determining the
density associated with a particular channel required evaluation using
classical trajectories.  The non-Hermitian character introduced by the
projection operator $\hat{R}_r$ onto channel $r$ allowed for the
successful reproduction of the entire complex term. We regard it as
particularly encouraging that use of classical trajectories in conjunction
with the complex Wigner transform of the term $|i\rangle \langle j|$
suffices to produce the imaginary part of the interference contribution
with such accuracy. We note, however, that the method is not expected to
produce resonance structures in the cross section, being most useful for
systems where the dynamics is direct and
short-lived\cite{Hupper,Hupper1,Miller2}.

Applications to full 3 dimensional photodissociation computations and to
scattering (for bimolecular coherent control\cite{cc,bmcc} are underway.

\begin{acknowledgments}
This work was supported by the U.S. Office of Naval Research and by Photonics
Research Ontario.
\end{acknowledgments}

\end{document}